\documentclass[preprint,aps,pra,notitlepage, amsmath,showpacs]{revtex4-1}

\usepackage[T1]{fontenc}\usepackage[latin1]{inputenc}
\usepackage[pdftex]{graphicx,color}
\usepackage{rotating}
\usepackage{bm} 
\usepackage{amssymb}
\usepackage{amsmath}
\usepackage{bm}
\usepackage[capitalize]{cleveref}
\usepackage{ulem}
							
\newcommand{\beq}{\begin{equation}}
\newcommand{\eeq}{\end{equation}}
\newcommand{\bra}[1]{\langle #1 |}
\newcommand{\ket}[1]{| #1 \rangle}

\newcommand{\half}{\frac{1}{2}}

\newcommand{\akl}{a_{\bm{k}\lambda}}

\newcommand{\beqs}{\begin{subequations}}
\newcommand{\eeqs}{\end{subequations}}

\newcommand{\nbar}{\bar{n}}

\begin{document}
\date{\today}
\flushbottom
\title{Secular vs. Non-secular Redfield Dynamics and Fano Coherences in Incoherent Excitation: An Experimental Proposal}
\author{Amro Dodin\footnote{Current address:  Department of Chemistry, MIT, Cambridge, Massachusetts  02139}}
\affiliation{Chemical Physics Theory Group, Department of Chemistry, University of Toronto, Toronto, Ontario, M5S 3H6, Canada}
\author{Timur Tscherbul}
\affiliation{Department of Physics, University of Nevada, Reno, Nevada 89557}
\author{Robert Alicki}
\affiliation{Faculty of Mathematics, Physics and Informatics, University of Gdansk, ul. Wita Stwosza 57, 80-952 Gdansk, Poland}
\author{Amar Vutha}
\affiliation{Department of Physics, University of Toronto, Toronto, Ontario M5S 1A7}
\author{Paul Brumer}
\affiliation{Chemical Physics Theory Group, Department of Chemistry, and Center for Quantum Information and Quantum Control, University of Toronto, Toronto, Ontario, M5S 3H6, Canada}

\begin{abstract}

Two different Master Equation approaches have been formally derived
to address
the dynamics of open quantum systems interacting with a thermal environment  (such as sunlight).
They have led to two different physical results:
non-secular equations that show noise-induced (Fano)
coherences and secular equations that do not. An experimental test for
the appearance
of non-secular terms is proposed using Ca atoms in magnetic fields  excited by broadband incoherent radiation. Significantly
different patterns of fluorescence are predicted, allowing for a clear
test for the validity of the  secular and non-secular approach and for the observation of Fano coherences.

\end{abstract}

\maketitle
\newpage

\section{Introduction}
\label{sec:Intro}

\label{sec:ME}

Quantum Master Equations (ME's) are an essential tool for the
study of the dynamics of open quantum systems, i.e. where a system
interacts with an unmonitored environment\cite{breuer_theory_2007}.
Formally exact integro-differential Master Equations are known
\cite{nakajima_quantum_1958, zwanzig_ensemble_1960}, but are
often intractable, reflecting
the complexity of the full quantum (system + bath) dynamics. As
a result, ME's are often simplified  via  a weak-coupling (Born-Markov)  approximation, and resulting in
two general classes of equations: secular Master
Equations where the system coherences and populations are uncoupled,
and non-secular equations where this coupling is manifest. From a mathematical
perspective, as discussed below, general arguments exist in favor of
both of these treatments, although they can give very different results \cite{alicki_quantum_2007}.
Convincing experimental tests to discern which of the two approaches is
physically correct are lacking and are sorely needed.

Recent studies have focused on the dynamics of quantum systems excited
by natural incoherent light such as thermal noise or sunlight,
often as a means of
understanding  natural  processes such as photosynthesis or vision.
Such
studies \cite{dorfman_photosynthetic_2013,scully_quantum_2011,kozlov_inducing_2006,hegerfeldt_coherence_1993,
agarwal_quantum_2001,agarwal_quantum_1974,tscherbul_partial_2015,tscherbul_long-lived_2014,dodin_quantum_2016,dodin_slow_2016-1},
 when   using non-secular Master Equations,
show significant system coherences, called Fano coherences. Here we propose
an experiment that would clearly expose the role of coherences in
natural light excitation and, in doing so, provide experimental tests
for the validity of either the secular or non-secular treatments of systems
excited by natural incoherent radiation in parameter regions where
the different Master Equations appear equally valid.

Fano coherences, although not as yet observed, have been suggested as important features in natural light harvesting \cite{tscherbul_long-lived_2014,dodin_quantum_2016} as well as significant in improving the efficiency of quantum
heat engines \cite{scully_quantum_2011}.  Hence, observing Fano coherences also serves
as motivation for the proposed experimental study.


\subsection{Secular and Non-secular Master Equations}

\label{subsectB}

Most common amongst the approximations used to simplify
exact Master Equations is
the Born-Markov approximation assuming weak system-bath coupling and
vanishing bath memory time \cite{breuer_theory_2007,valkunas_book_2013}.
However,
improperly applied, these approximations can lead to unphysical
behavior, such as negative state populations
or sensitivity of the dynamics to a non-interacting spectator
system \cite{alicki_quantum_2007,creatore_efficient_2013}.
For example, the non-secular Redfield equations that result
from a na\"{i}ve application of the Born-Markov approximation
do not guarantee the preservation of positive populations
\cite{alicki_quantum_2007}, and such negative
populations, as well as diverging populations, have been
shown to plague simulations of a suggested
coherence-enhanced heat engine \cite{creatore_efficient_2013}.

Requiring ``complete positivity'' provides a rigorous
 condition for physically meaningful dynamics
 \cite{alicki_quantum_2007}. In this case, any reduced dynamics with initial uncorrelated system-bath state is completely positive and preserves positivity in the presence of entangled non-interacting spectator systems. The general form of
 completely positive MEs has been obtained
 \cite{kossakowski_quantum_1972,lindblad_generators_1976,gorini_completely_1976}
 for discrete systems and are often referred to as Lindblad
 equations, or are said to be of Lindblad form
 \cite{breuer_theory_2007}. Ultimately, the exact ME
 governing the dynamics of a real system must be of Lindblad
 form, although obtaining and solving it is generally quite
 difficult.

Unfortunately, a given Lindblad equation may not correctly
model the physics of the system. For example, applying
the standard secular approximation to the Redfield
equation
decouples the coherences from the populations, removes
rapidly oscillatory terms, and yields the
secular  Redfield  equations,  a Lindblad Master Equation
\cite{breuer_theory_2007} with guaranteed positive populations.
However, if not judiciously applied, this approximation can
neglect significant contributions to the system dynamics. For
example, in the case of a system driven by incoherent light (as
is of interest in this paper), the secular Redfield equations
can miss interference effects that appear in non-secular or
partial  secular  treatments
\cite{dorfman_photosynthetic_2013,scully_quantum_2011,kozlov_inducing_2006,hegerfeldt_coherence_1993,agarwal_quantum_2001,agarwal_quantum_1974,tscherbul_partial_2015,tscherbul_long-lived_2014,dodin_quantum_2016,dodin_slow_2016-1}.
Hence, in essence,
non-secular Redfield equations may capture the underlying
physics more effectively than their secular form in their
domain  of  applicability,  but  they  can present basic
mathematical problems insofar as they may be non-positive.
Despite these fundamental issues,
Redfield theory continues to be widely used due to the physical
intuition provided by the perturbative approach upon which it is
based.

These concerns suggest returning to the formal derivations
of the Redfield Equations, which include conditions on their
validity.


\subsection{Approximations and the Davies limits}

In this regard, Davies has provided a rigorous
derivation  of  a secular Redfield Equation
\cite{davies_markovian_1974} in the weak coupling limit that
retains the physical
intuition  of  the  perturbative  approach, guarantees
completely  positive dynamics, gives rigorous
conditions  for the Born, Markov and secular approximations, and
defines the domain where the resultant equations apply. Formally,
the  Davies  weak  coupling  limit  is  correct in the limit of
vanishing  system-bath  coupling constant $\lambda$, but can be
applied  in  practice  when  the  system  relaxation,
$\tau\propto\lambda^{-2}$, is much slower than any oscillations
in the system. Since the oscillation frequencies are determined
by  the energy spacing of non-degenerate system eigenstates, this implies that
$\tau\propto\lambda^{-2}\gg \omega^{-1}$ where $\hbar\omega$ is
the smallest non-zero energy  difference  in  the  system.
Interestingly,  this approximation also holds for systems with exactly degenerate eigenstates.
If  these conditions are not satisfied then the secular approximation fails and the
resultant  equations fail to model the system dynamics.
Under  these  conditions,  some version of non-secular dynamics
may be required.

Significantly,  Davies  generalized  the weak coupling limit to
treat  systems  with  ``nearly  degenerate''  states  where the
secular  approximation does not apply \cite{davies_model_1978}.
This  is  done by first assuming that the ``nearly degenerate''
states  are  degenerate  and  then applying the original Davies
limit  to the resulting approximate system.
The  energy  difference between these states is then introduced
as  a  perturbation  to  remove  the degeneracy. The resulting
Master Equation  retains  the  non-secular  terms  and  hence
non-secular effects between  nearby  states.  However,  the
perturbative  correction  that  introduces  the energy shift is
only accurate  for  small  energy  splitting  between  nearly
degenerate  states.  This leads to two ``complementary'' completely
positive  approximations,  the  secular  approximation  and the
non-secular  perturbative  energy  shift:  two  equations  with
dramatically different dynamics, particularly in the overlapping
region where  $\tau\propto\lambda^{-2}\approx \omega^{-1}$. Indeed,
the dynamics in the regime might well be non-Markovian \cite{Alicki1989}.


The  distinction  between secular and non-secular QME's and the
formal  method for obtaining them  has  become
increasingly  relevant in the study of
quantum systems driven by incoherent light. In this case the
secular  treatment  neglects interference effects and
gives  rate-law  equations  that  reproduce
Einstein's theory of light-matter interaction.
By  contrast, the non-secular treatment
retains the interference effects  and  produces  markedly
different  coherent  dynamics  before approaching the
rate-law-predicted steady state
\cite{dorfman_photosynthetic_2013,scully_quantum_2011,kozlov_inducing_2006,hegerfeldt_coherence_1993,agarwal_quantum_2001,agarwal_quantum_1974,tscherbul_partial_2015,tscherbul_long-lived_2014,dodin_quantum_2016,dodin_slow_2016-1}.
Computational examples of the latter include fluorescence
in  V  and  $\Lambda$  systems \cite{hegerfeldt_coherence_1993}
with associated non-secular-based effects on the time resolved
fluorescence  of  systems  with  closely spaced states,
the prediction \cite{kozlov_inducing_2006} of a
population-locked  state  in  V-systems with degenerate excited
states  pumped by a single incoherent field,  and a  heat  engine
with  enhanced  power due to noise-induced coherences
\cite{scully_quantum_2011}.  We  have  recently shown,  in
theoretical  studies,  long-lived  quasistationary coherences in V-systems
with  nearly  degenerate excited states
\cite{dodin_quantum_2016,dodin_slow_2016-1,tscherbul_partial_2015,tscherbul_long-lived_2014}.
In addition, the proper
approach  of  these  systems  to  an  incoherent  thermodynamic
equilibrium state has  been  discussed in detail \cite{agarwal_quantum_2001}.


Although significant non-secular
effects of this kind have been predicted in theoretical studies
of noise-induced coherences  \cite{agarwal_quantum_2001,agarwal_quantum_1974,tscherbul_partial_2015,tscherbul_long-lived_2014,dodin_quantum_2016,dodin_slow_2016-1}  they have not been verified
experimentally.  Specifically, these Fano coherences have not been experimentally observed.   Furthermore, as noted above,
theoretical issues remain unresolved. Here we propose an
experiment to address these
fundamental theoretical issues by
examining the role of non-secular contributions to the dynamics
of incoherently excited quantum systems over a large parameter range.
In doing so we would
determine whether the secular or non-secular QME is the
physically appropriate version of the Master Equations derived
by Davies. In the proposed experiment the distinction is clear:
the secular result gives no contribution from coherences,
whereas the non-secular coherence contribution is either
constant in time or oscillatory, depending on system
parameters. Scanning the experimental parameters allow for the
study of a full range of behaviors.

Below, Sect.\ref{sec:Sys} describes the Calcium system proposed for experimental study.  Section \ref{sectIII} derives dynamical results for this system in both secular and non-secular descriptions.  Predicted experimental signals are shown in Sect. \ref{sec:Detect}.

\section{Proposed System}
\label{sec:Sys}
\begin{figure}[htp] 
\includegraphics[width=\textwidth]{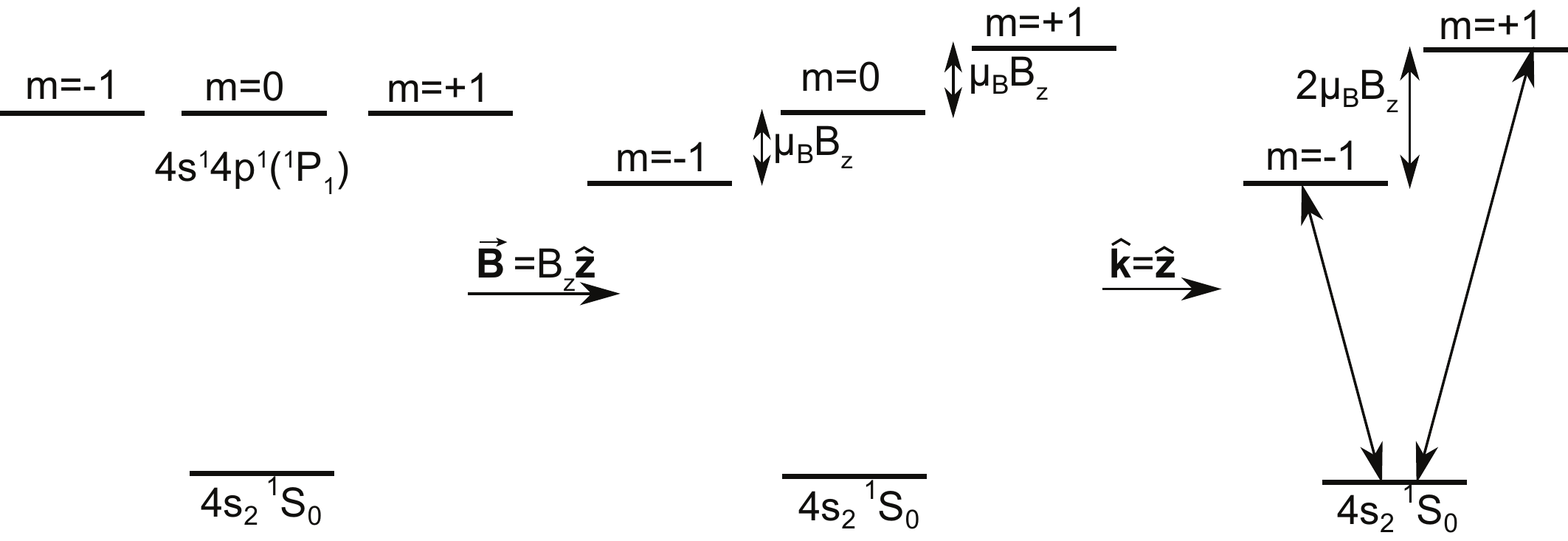}
\caption{Sketch of the V-subsystem of Calcium excited in proposed experiment. The magnetic field, $\bm{B}$ of magnitude $B$, and wavevector of incident light, $\bm{k}$, are parallel along the $\hat{z}$ direction. The excited state (Zeeman) splitting is given by $2\mu_BB$ where $\mu_B$ is the magnetic dipole moment of the $4p^\pm$ states. The rightmost figure shows the $m$=-1 and $m$=+1 levels,
denoted $|e_1\rangle$ and $|e_2 \rangle$ below.}
\label{fig:Model}
\end{figure}

To examine the role of non-secular contributions to system
dynamics we propose a V-system with tunable excited state
splitting. In particular we focus on atomic $s \rightarrow p$ transitions
with only two of the \textit{p}-state angular momentum $m$-sublevels
 excited. This is achieved by irradiating an atom with a
beam of incoherent light propagating along  the $\hat{z}$ direction that
excites the orthogonal $p$ states in the x-y plane. By applying
a magnetic field parallel to the incident light beam, a tunable
Zeeman shift can be used to study the dynamics as a function of
the spacing between the $p_\pm$ energy levels. However, due to
the orthogonal polarization (circularly polarized $\sigma^\pm$)
of $s\rightarrow p_\pm$ transitions, an unpolarized incoherent light beam will
produce identical dynamics from both the secular and
non-secular Master Equations. To distinguish these cases, a
beam of spectrally broadened light polarized in the $\hat{x}$
direction can be used.  This scenario can be experimentally
realized with a beam of Calcium atoms excited by a
polarized spectrally broadened laser in a uniform magnetic field. The
resultant V-system generated between the doubly occupied
singlet ground state ($4s_2$ $^1S_0$) denoted $|g\rangle$ and the excited triplet
states ($4s_14p_1$ $^1P_1$; $m_j=\pm1$) denoted $|e_i \rangle$ is shown schematically
in \cref{fig:Model}.
The incoherent beam is assumed sufficiently weak that the
population of doubly excited $4p_2$ states can be neglected.

As shown below, incoherent excitation from the calcium $4s_2^1S_0$ ground state will generate
coherences between the excited $m_j = \pm1$ triplet states $4s{_1}4{p_1}{ ^1}P{_1}$ if non-secular
terms contribute. No such coherences will be generated from secular terms. As such, observing
coherences, e.g., by detecting quantum beats in the spontaneous emission from the Ca atoms, proves
the presence of non-secular contributions to the Master Equation.

A summary of the experimentally accessible range of parameters is given in Table I, including the excited state linewidth $\gamma$ and $\tau_\gamma = \gamma^{-1}\approx T_{transit}/4000$, where $T_{transit}$ is the time over which the atom encounters the broadband laser beam.
Furthermore, the bandwidth of the spectrally broadened laser is $\Delta\nu$, larger than the largest excited-state splitting $\Delta$, i.e., $\Delta\nu > 2\max{\Delta}$, justifying the Wigner-Weisskopff approximation used below.
The range of accessible $\Delta$ values spans both limiting cases of $\Delta \gg \gamma$ and $\Delta \ll \gamma$. Indeed, as one scans from large $\Delta/\gamma$ to small $\Delta/\gamma$ one expects to observe a
transition from the Davies weak coupling regime (secular) to the Davies perturbative regime (non-secular).
In addition, this approach probes the intermediate regime ($\Delta \sim \gamma$) where neither of the rigorously derived Master Equations hold.

\begin{table}
\begin{tabular}{l|l}
Parameter & Magnitude \\ \hline \hline
Linewidth of Excited states ($\gamma$) & $2\pi \times 34.6$ MHz \\
Ground-Excited state transition frequency ($\omega_0$) & $709.1$ THz \\
Excited State Splitting ($\Delta$) & 0 to $2\pi\times400$ MHz\\
Transition Dipole Moment ($\mu_\pm$) & 2.85 $e$ $a_0$ \\
Light-Atom Interaction time ($T_{transit}$) & 20 $\mu$s \\
Broadband Laser Output Power ($P_{laser}$) & 20 mW \\
Laser Spectral Width ($\Delta\nu$) & $2\pi \times 1$ GHz\\
\end{tabular}
\caption{Summary of experimental parameters for Calcium V-system shown in
\cref{fig:Model}. }
\label{tab:Param}
\end{table}

\section{Theoretical Predictions}
\label{sectIII}
We provide below a fundamental derivation, directly from the Davies approach, of the ME's associated with this scenario.
An alternative derivation, to connect with earlier results \cite{tscherbul_partial_2015,tscherbul_long-lived_2014,dodin_quantum_2016,dodin_slow_2016-1}, is presented in the Appendix.

\subsection{Completely positive Master Equation for V-system}
Consider first the most general phenomenological Master Equation obtained from the Lindblad form which takes into account only transitions in the $V$-system and
neglects leaking of probability and pure dephasing. We write down the equations for $\rho_{e_i e_i}$, and  $\rho_{e_1 e_2} $  only and omit the Hamiltonian contribution because it depends on the choice of the excited state basis (we  keep the freedom of choosing convenient basis in the corresponding 2-dimensional Hilbert subspace)
\begin{equation}
\dot{\rho}_{e_i e_i} = R_{ii} \rho_{gg} - K_{ii} \rho_{e_i e_i} - \frac{1}{2} K_{12} \rho_{e_1 e_2} - \frac{1}{2} K_{21}\rho_{e_2 e_1}
\label{MEaa}
\end{equation}
\begin{equation}
\dot{\rho}_{e_1 e_2} =  R_{12} \rho_{gg}- \frac{1}{2}(K_{11} + K_{22})\rho_{e_1 e_2} - \frac{1}{2} K_{12}(\rho_{e_1 e_1} + \rho_{e_2 e_2} )
\label{MEab}
\end{equation}
The requirement of complete positivity  implies that  $[R_{ij}]$ and $[K_{ij}]$ are positive definite, i.e.,
\begin{equation}
R_{ii}, K_{ii} \geq 0 , \quad  |R_{12}|^2 \leq R_{11} R_{22},\quad |K_{12}|^2 \leq K_{11} K_{22} ,
\label{CP}.
\end{equation}
The functional form of $R_{ij}$ and $K_{ij}$ for the Calcium system is provided later below.

\subsection{Master Equations from Davies' Approach}
The interaction of a V-system with radiation is given in the dipole approximation by
\begin{equation}
H_{int} = - \mathbf{D}\cdot\mathbf{E}_{reg}
\label{H_int}
\end{equation}
where $\mathbf{D}$ is an atomic dipole operator which can be defined in terms of the excited state basis $\ket{e_1}=\ket{4p_x}$ and
$\ket{e_2} = \ket{4p_y} $
\begin{equation}
\mathbf{D} = (\ket{g}\bra{e_1} + \ket{e_1}\bra{g})\bm{\mu}_{g e_1} + (\ket{g}\bra{e_2} + \ket{e_2}\bra{g})\bm{\mu}_{g e_2} .
\end{equation}
Here, $\bm{\mu}_{g e_i}$ is the corresponding transition dipole moment and
$\mathbf{E}_{reg}$ is the electric field ($\nu$ - polarization)
\begin{equation}
\mathbf{E}_{reg} = i \sum_{\nu =1}^2 \sum_{|\bm{k}| \leq K} \Bigl(\frac{\hbar c|\bm{k}|}{2\epsilon_0 V}\Bigr)^{1/2}  \bm{\epsilon}_{\bm{k},\nu}
\Bigl\{a_{\bm{k},\nu} - {a^{\dagger}}_{\bm{k},\nu}\Bigr\},
\label{E_reg}
\end{equation}
where $\bm{k}$ is the wavevector, and $\nu$ denotes the polarization, $a_{\bm{k},\nu}$ and $a^{\dagger}_{\bm{k},\nu}$ are the annihilation and creation operators of the mode with wavevector $\bm{k}$
and polarization $\nu$.

Here the incoherent radiation is assumed to be described by a stationary state with the photon population numbers  $n(\bm{k},\nu)$ defined by (for $V\to\infty$)
\begin{equation}
\langle {a^{\dagger}}_{\bm{k},\nu}a_{\bm{k}',\nu'}\rangle = n(\bm{k},\nu) \delta (\bm{k}-\bm{k}') \delta_{\nu\nu'}.
\label{pop}
\end{equation}
We consider the two cases of ME resulting from Davies' procedure below.\\

The derivations of ME using the Davies weak coupling limit  combines in a single limiting procedure "Born" , "Markovian" and "secular" approximations. The basic ingredient is the transition to the interaction picture followed by the suitable averaging  of oscillating terms. The averaging process depends on the relevant time-scales and creates some ambiguity. Consider a system Hamiltonian of the form  $H = H_0 + \epsilon V$, where $H_0$ possesses degeneracies and where $V$  removes at least some of them. Then if $\epsilon$ is "small", i.e., the level splitting generated by $V$ is small in comparison with typical relaxation rates, we can switch to the interaction picture,
derive the ME with $H_0$, and subsequently introduce $\epsilon V$
as a perturbation. In the opposite case of large splitting we must use the full Hamiltonian $H$ in the interaction picture. Both regimes lead to different forms of ME, called here non-secular and secular respectively. The crossover regime is non-Markovian and needs special treatment. Similar phenomena are not uncommon in quantum mechanics, with a notable example of LS-coupling versus JJ-coupling in atomic physics, where the intermediate angular momentum region needs to be treated separately.

\par

\subsubsection{Non-secular ME}

Consider the case of small Zeeman splitting $\Delta \ll r$, where $r$ is the pumping rate from the ground state to   $\ket{e_1}$  and  $\ket{e_2}$.  Here     we apply the interaction picture and the derivation of the Master Equation with
\begin{equation}
H_0 = \hbar\omega_0 \bigl(\ket{e_1}\bra{e_1} + \ket{e_2}\bra{e_2}), \quad \omega_0 = \frac{1}{2}(\omega_1 + \omega_2)
\label{ham0}
\end{equation}
and add the splitting term at the end.
\par
Then the standard computation yields
\begin{equation}
K_{ij} = \Gamma_{ij} +R_{ij}
\label{K_matrix}
\end{equation}
\begin{equation}
\Gamma_{ij} = \frac{\pi c}{\hbar \epsilon_0}\sum_{\nu =1}^2 \int d^3\bm{k} |\bm{k}|(\bm{\mu}_{g e_i} \cdot \bm{\epsilon}_{\bm{k},\nu} )(\bm{\mu}_{g e_j} \cdot \bm{\epsilon}_{\bm{k},\nu} )\delta (|\bm{k}|-\omega_0)
\label{gamma_matrix}
\end{equation}
\begin{equation}
R_{ij} = \frac{\pi c}{\hbar \epsilon_0}\sum_{\nu =1}^2 \int d^3\bm{k} |\bm{k}|n(\bm{k},\nu) (\bm{\mu}_{g e_i} \cdot \bm{\epsilon}_{\bm{k},\nu} )(\bm{\mu}_{g e_j}\cdot \bm{\epsilon}_{\bm{k},\nu} )\delta (|\bm{k}|-\omega_0)
\label{R_matrix}
\end{equation}
with all three matrices  positively defined. From the structure of the V-system it follows that $\bm{\mu}_{g e_1}\bot\bm{\mu}_{g e_2}$ and $|\bm{\mu}_{g e_1}|=|\bm{\mu}_{g e_2}|$,  which implies that in the space of $|e_1\rangle$ and $|e_2\rangle$
\begin{equation}
\Gamma =\left[
\begin{array} {cc}
\gamma &  0 \\
0 & \gamma  \\
\end{array}
\right] , \quad \gamma = \frac{|\bm{\mu}_{g e_i}|^2 \omega_0^3}{3\pi \epsilon_0 c^3}
\label{eq:Gamma}
\end{equation}
Pumping provided by a spectrally broadened laser beam polarized along $x$-direction leads to the following form for the matrix $R$:
\begin{equation}
R =\left[
\begin{array} {cc}
2r &  0 \\
0 & 0  \\
\end{array}
\right] .
\label{eq:R}
\end{equation}
In the next step we transform the Master Equation into the new basis of excited states which corresponds to eigenvectors of the total Hamiltonian including Zeeman splitting. Namely, now
\begin{equation}
\ket{e_1} =   \frac{1}{\sqrt{2}}\bigl(\ket{4p_x} + i\ket{4p_y}\bigr), \quad \ket{e_2} =   \frac{1}{\sqrt{2}}\bigl(\ket{4p_x} -i \ket{4p_y}\bigr)
\label{newbasis}
\end{equation}
and in this new basis the matrices  $\Gamma$ and $R$ have form
\begin{equation}
\Gamma =\left[
\begin{array} {cc}
\gamma  &  0  \\
0 & \gamma   \\
\end{array}
\right] ,
R = \left[
\begin{array} {cc}
r &  r \\
r & r  \\
\end{array}
\right]
\label{eq:newR}
\end{equation}
Adding the splitting Hamiltonian  $V = \frac{1}{2} \hbar\Delta \bigl(\ket{e_1}\bra{e_1} - \ket{e_2}\bra{e_2})$ one obtains the ME for the relevant matrix elements
\begin{eqnarray}
\dot{\rho}_{e_i e_i} & = & r \rho_{gg} - (\gamma + r) \rho_{e_i e_i} - \frac{r}{2} (\rho_{e_1 e_2} +\rho_{e_2 e_1}) \nonumber \\
\dot{\rho}_{e_1 e_2}& = &  r\rho_{gg} - (\gamma + r) \rho_{e_1 e_2}  -i\Delta \rho_{e_1 e_2}  - \frac{r}{2}(\rho_{e_1 e_1}+ \rho_{e_2 e_2})
\label{newnonsec}
\end{eqnarray}

It is important to note that, in contrast with the Master
Equation   associated   with   isotropic   radiation
(see Appendix and Refs. \cite{tscherbul_partial_2015,tscherbul_long-lived_2014,dodin_quantum_2016,dodin_slow_2016-1}) the coefficient of the last term
in these equations for $\dot{\rho}_{e_i e_i}$ and for $\dot{\rho}_{e_1 e_2}$
is $- \frac{1}{2} r $, as opposed to $- \frac{1}{2}(r+\gamma)$. This is
a consequence of the Ca atom being coupled to two different photon baths,
the isotropic radiative environment and the directional excitation beam.
As a result, and as shown below, long lived coherences can survive in the
non-secular case.

\subsubsection{Secular ME}
Consider the case of large Zeeman splitting $\Delta \gg r $ . Then we apply the interaction picture and the derivation of ME with the full Hamiltonian including the splitting term
\begin{equation}
H = H_0 +\frac{1}{2}\hbar\Delta \bigl(\ket{e_1}\bra{e_1} - \ket{e_2}\bra{e_2}) .
\label{ham1}
\end{equation}
Then, in principle, the diagonal elements of the  $\Gamma$ - matrix in Eq. \eqref{eq:Gamma} are different, because they are computed using Eq. \eqref{gamma_matrix} at two different frequencies.  However, for simplicity we assume that they are equal. This is equivalent to the Wigner-Weisskopf approximation often used in quantum optics \cite{scully_quantum_1997}.
In this case the off-diagonal matrix elements in Eq. \eqref{eq:newR} dissappear and the ME equation  simplifies to
%
%
%
\begin{eqnarray}
\dot{\rho}_{e_i e_i} & = &  r \rho_{gg} - (\gamma + r) \rho_{e_i e_i} \nonumber \\
\dot{\rho}_{e_1 e_2} & = & - (\gamma + r) \rho_{e_1 e_2}  -i\Delta \rho_{e_1 e_2}.
\label{newsec}
\end{eqnarray}

\subsection{Solutions}

Consider the case of weak incoherent pumping by a beam of uniform intensity and well-defined beam area.
An incident beam of ground state  Calcium atoms would experience the light as a suddenly turned-on field
at $t=0$, the time that it enters the incoherent beam.
For simplicity consider the case where the measurement of atomic polarization is conducted in the interaction region,
so that the atom experiences a constant field intensity between $t=0$ and the time of measurement.
Given the normalization condition $\rho_{gg}+\rho_{e_1e_1}+\rho_{e_2e_2} = 1$ and using the notation $\bm{\rho}=\left[\rho_{e_1e_1},\rho_{e_2e_2},\rho_{e_1e_2}^{Re},\rho_{e_1e_2}^I\right]$,  where $\rho^{Re}_{e_1 e_2}$ and $\rho_{e_1e_2}^I$   are the
real and imaginary parts of $\rho_{e_1e_2}$.   Equations (\ref{newnonsec}) and (\ref{newsec})             for the evolution of the V-system can be written in vector form as:
\begin{equation}
\dot{\bm{\rho}}=A\bm\rho+\bm{d}
\label{eq:VecQME}
\end{equation}
where the secular and non-secular equations differ in their coefficient matrices, $A$ and driving vectors $\textbf{d}$.
In \cref{eq:VecQME} the dynamics of the excited states are naturally divided into two parts, the pumping from the ground state given by $\bm{d}$ and the "internal" dynamics of the excited manifold contained in $A$.
The non-secular evolution  Eq. (\ref{newnonsec}) is characterized by
\begin{subequations}
\label{eq:NSVec}
\begin{equation}
A_{\textrm{NS}}=\left[
\begin{array} {cccc}
-\gamma-2{r} & -r & -r & 0 \\
-r & -\gamma-2{r} & -r & 0 \\
-\frac{r}{2} & -\frac{r}{2} & -\gamma -r & \Delta \\
0 & 0 & -\Delta & -\gamma-r
\end{array}
\right]
\label{eq:NSA}
\end{equation}
\begin{equation}
\bm{d}_{\textrm{NS}} = \left[
\begin{array}{c}
r \\ r \\ r \\ 0
\end{array}
\right]
\label{eq:NSd}
\end{equation}
\end{subequations}
While the secular evolution [Eq. (\ref{newsec}) ] is characterized by an absence of population-coherence coupling, with
\begin{subequations}
\label{eq:SVec}
\begin{equation}
A_{\textrm{S}}=\left[
\begin{array} {cccc}
-\gamma-2r & -r & 0 & 0 \\
-r & -\gamma-2r & 0 & 0 \\
0 & 0 & -\gamma -r & \Delta \\
0 & 0 & -\Delta & -\gamma-r
\end{array}
\right]
\label{eq:SA}
\end{equation}
\begin{equation}
\bm{d}_{\textrm{S}}=\left[
\begin{array}{c}
r \\ r \\ 0 \\ 0
\end{array}
\right]
\label{eq:Sd}
\end{equation}
\end{subequations}

Essential differences between the secular and non-secular equations are readily seen in these expressions.
That is, the driving vectors [\cref{eq:NSd,eq:Sd}] clearly show that in the non-secular dynamics the field will drive ground state Calcium atoms to coherent superpositions of the excited states, whereas the secular equations predict that the system will be driven to an incoherent mixture of excited states \cite{dodin_quantum_2016}.

In general, the solution to \cref{eq:VecQME} is given by \cite{boyce_elementary_2009}
\beq
\bm\rho = \int^t_0 ds e^{A(t-s)}\bm{d} \to \sum_{i=1}^4 \int^t_0 ds (\bm{v_i}\cdot\bm{d})e^{\lambda_i (t-s)} \bm{v_i}
\label{eq:GenSol}
\eeq
where $\lambda_i$ is the $i^{\textrm{th}}$ eigenvalue of the coefficient matrix $A$ and $\bm{v_i}$ is the corresponding eigenvector.
Hence, upon solving for the spectral decomposition of $A$ the evolution equation is reduced to a simple exponential integral.

For the non-secular coefficient matrix \cref{eq:NSA}, the characteristic polynomial is given by:
\begin{equation}
\textrm{CharPoly}(A_{\textrm{NS}})= (\lambda+\gamma+r)\left[(\lambda+\gamma+r)^3+2r(\lambda+\gamma+r)^2+(\Delta^2-r^2)(\lambda+\gamma+r)+2r\Delta^2 \right]~.
\label{eq:CharPolyNS}
\end{equation}
 It is seen to be comprised of a term linear in $\lambda$ giving the total decay mode $\lambda_4=-(\gamma+r)$, approximated by $\lambda_4 \approx -\gamma$ in the weak pumping limit ($\nbar =2r/\gamma \ll1$), and a cubic term giving the remaining normal modes.

The contribution of spontaneous emission in \cref{eq:CharPolyNS,eq:CharPolyS} is contained entirely in the term $(\lambda+\gamma+r)$.
Physically, this implies that the only influence of spontaneous emission is to introduce a uniform decay to all normal modes of the system.
 To obtain more explicit expressions for the evolution of the system under the non-secular dynamics we consider two complimentary limits below, those of large and small Zeeman splitting relative to the incoherent pumping rates (i.e., $r\gg \Delta$ and $r\ll\Delta$).

The secular coefficient matrix \cref{eq:SA} has a simpler polynomial that can be factorized into the following biquadratic form:
\begin{equation}
\textrm{CharPoly}(A_{\textrm{S}})=\left[(\lambda+\gamma+r)^2+\Delta^2\right]\left[(\lambda+\gamma+r)^2+r^2\right]
\label{eq:CharPolyS}
\end{equation}
It will be shown below that these terms correspond to two damped oscillatory modes corresponding to the evolution of the coherences ($\lambda^{(\textrm{S})}_{3,4}=-(\gamma+r)\pm i\Delta\approx-\gamma \pm i \Delta$) and two exponential modes corresponding to the population ($\lambda^{(\textrm{S})}_{1,2}=-(\gamma+r)\pm r\approx -\gamma$).

One note is in order. Our solutions assume the sudden turn-on of
the interaction
between Ca atoms and the incoherent radiation. Since slow turn-on relative to
system dynamical time scales is known \cite{amrslow,pachonjphysb} to significantly reduce
the magnitude of any induced coherences, we note that sufficiently rapid
turn-on, using acousto-optic modulators, is possible in these systems.

\subsubsection{Large Zeeman Splitting $\Delta/r \gg1$}
\label{subsec:BigDel}
In the limit of large Zeeman Splitting the rate at which coherences oscillate is much faster than the rate $r\rho_{e{_1}e{_2}}$ of population to coherence coupling. A binomial approximation to lowest order in $r/\Delta \ll 1$ of the roots of the cubic term in \cref{eq:NSA} gives the eigenvalues of $A_\textrm{NS}$ in the  large splitting limit
\begin{subequations}
\label{eq:BNSLam}
\begin{equation}
\lambda_{1,2}\approx -\gamma
\end{equation}
\label{eq:BNSL12}
\begin{equation}
\lambda_{3,4}\approx -\gamma \pm i\Delta
\end{equation}
\label{eq:BNSL34}
\end{subequations}

Similarly, one can obtain the $\lambda$ eigenvalues for the secular case, which are found to be the same as the non-secular results.
The eigenvalues for the secular and non-secular equations agree
since the population to coherence coupling terms in \cref{eq:NSA} are small relative to the other matrix elements.
That is, since $r \ll \gamma \ll \Delta$, the dynamics of coherence to population coupling driven by the incoherent field is much slower than both the spontaneous emission and phase oscillation dynamics.
Hence, the coherence-population coupling terms do not contribute significantly to the evolution of the Calcium population dynamics.
The  secular and non-secular eigenvectors in this limit are also found to coincide  and are given by
\begin{subequations}
\label{eq:BV}
\beq
\bm{v}_1=[1,0,0,0]^T
\label{eq:BV1}
\eeq
\beq
\bm{v}_2=[0,1,0,0]^T
\label{eq:BV2}
\eeq
\beq
\bm{v}_{3,4}=[0,0,1,\pm i]^T
\label{eq:BV34}
\eeq
\end{subequations}
Hence, any difference between the secular and non-secular evolution in this  overdamped    $\Delta/r \gg 1$ regime can  be attributed to the treatment of the incoherent driving, i.e., \cref{eq:NSd} vs. \cref{eq:Sd}.

\begin{figure}[htp] 
\includegraphics[width=0.7\textwidth]{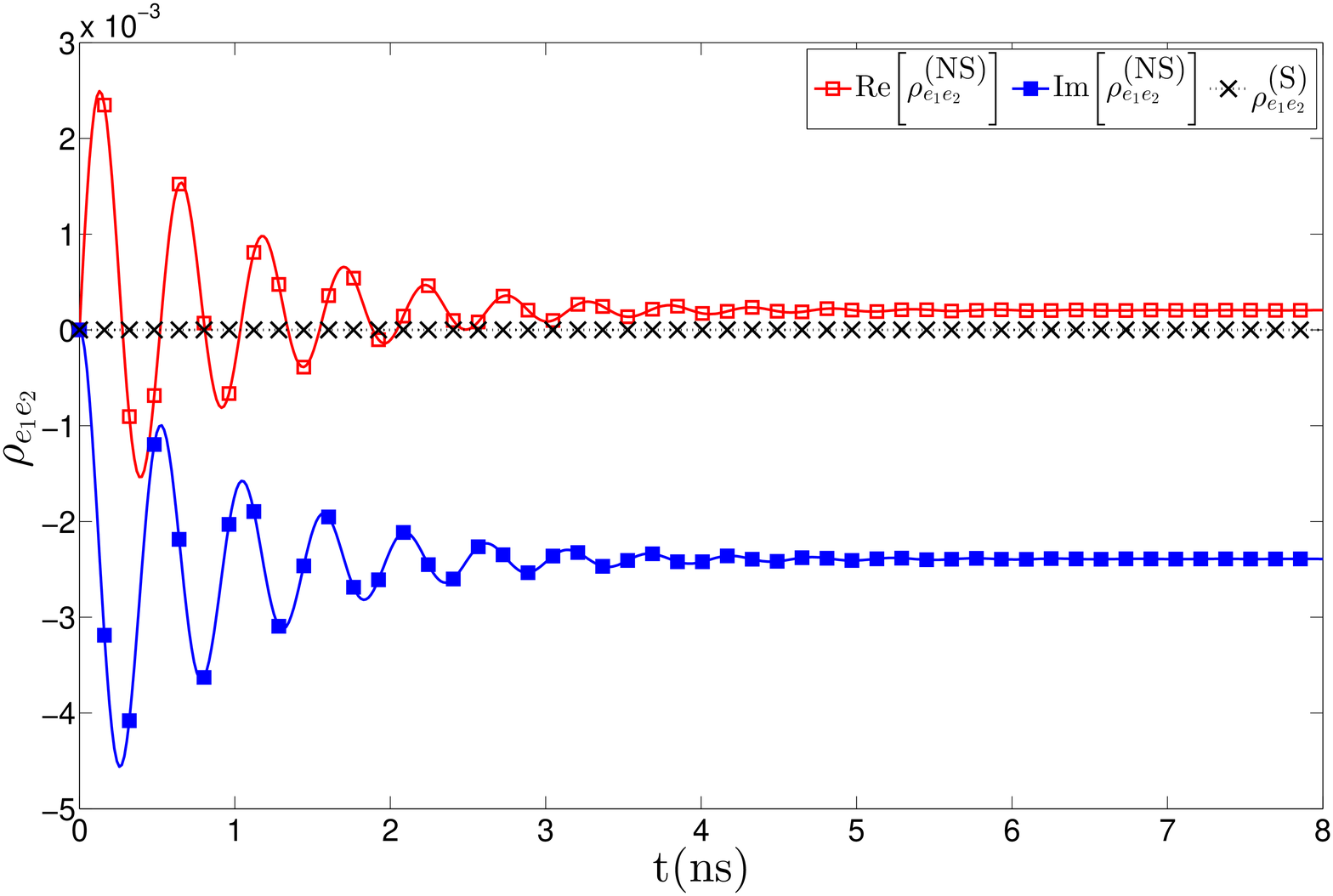}
\caption{Coherences of a Calcium atom in a sample large splitting regime $\Delta=12\gamma$ irradiated by a blackbody at $T=5800K$ whose average photon occupation number at the transition energies is $\bar{n}=0.0633$. The non-secular dynamics generate
oscillatory coherences from an initially incoherent ground state that survive for a time $\tau_\gamma = \gamma^{-1}$ while the secular evolution produces an incoherent mixture at all times.}
\label{fig:LargeSplit}
\end{figure}

Given these results, the secular dynamics are described by:
\begin{subequations}
\label{eq:BSDyn}
\beq
\rho_{e_ie_i}(t)=\frac{r}{\gamma}\left(1-e^{-\gamma t}\right)
\label{eq:BPop}
\eeq
\beq
\rho^{(\textrm{S})}_{e_1e_2}(t)=0
\label{eq:BSCoh}
\eeq
\end{subequations}
The non-secular population dynamics are the same [\cref{eq:BPop}], but the non-secular results show nontrivial coherence dynamics:
\beq
\rho^{(\textrm{NS})}_{e_1e_2}(t)=\frac{r}{\Delta}[e^{-\gamma t}\sin(\Delta t)+i\left(1-e^{-\gamma t}\cos(\Delta t)\right)]
\label{eq:BNSCoh}
\eeq
Hence, the secular dynamics predict that a statistical mixture of excited states will be produced while the non-secular dynamics predict a small but non-zero oscillating coherent component of the mixture as well as stationary imaginary coherences.
The coherences for both Master Equations in this regime, calculated exactly, are plotted in \cref{fig:LargeSplit}, where
 the distinction between the secular and non-secular evolution is evident. 
 (These exact results deviate slightly from 
 Eq. (\ref{eq:BNSCoh}) due to the inclusion of higher order terms).

The observed stationary coherences may at first appear to violate thermodynamics, since a system in equilibrium with a large bath is expected to be an incoherent mixture of energy eigenstates.
However, this is not the case for a system interacting with multiple baths.
Specifically, due to the anisotropy of the radiation field one can split the interaction into two dissipative baths.
The field modes with wavevector $\bm{k} || \hat{\bm{z}}$ and polarization $\hat{\bm{x}}$ behave as a hot bath, pumping energy into the Calcium V-system, while the remaining (vacuum) modes of the field behave as a cold bath to which energy is dissipated.
In this picture the excitation of Calcium by a polarized beam is a transport or heat engine problem between the hot directional field modes of the beam and the cold isotropic vacuum modes of the field.
Stationary coherences in systems interacting with two baths is not atypical, as the system operates out of equilibrium.

\subsubsection{Small Zeeman Splitting $\Delta/r \ll1$}
\label{subsec:SmallDel}
In the limit of a very weak magnetic field, and hence small $\Delta$, spontaneous emission will be the dominant influence on the undriven dynamics of the V-system. Accordingly, in both the secular and non-secular Master Equations, the four eigenvalues are approximately degenerate at
\beq
\lambda=-\gamma~,
\label{eq:SL}
\eeq
and the four-fold degenerate eigenvectors are given by $\bm{v}_i=\rho_i$ where $\bm{\rho}=[\rho_{e_1e_1},\rho_{e_2e_2},\rho_{e_1e_2}^{Re},\rho_{e_1e_2}^I]^T$.
Substituting this into \cref{eq:GenSol} gives the following secular dynamics:
\begin{subequations}
\label{eq:SSDyn}
\beq
\rho_{e_ie_i}(t)=\frac{r}{\gamma}\left(1-e^{-\gamma t}\right)
\label{eq:SSPop}
\eeq
\beq
\rho_{e_1e_2}^{(\textrm{S})}(t)=0~.
\label{eq:SSCoh}
\eeq
\end{subequations}

\begin{figure}[htp] 
\includegraphics[width=0.7\textwidth]{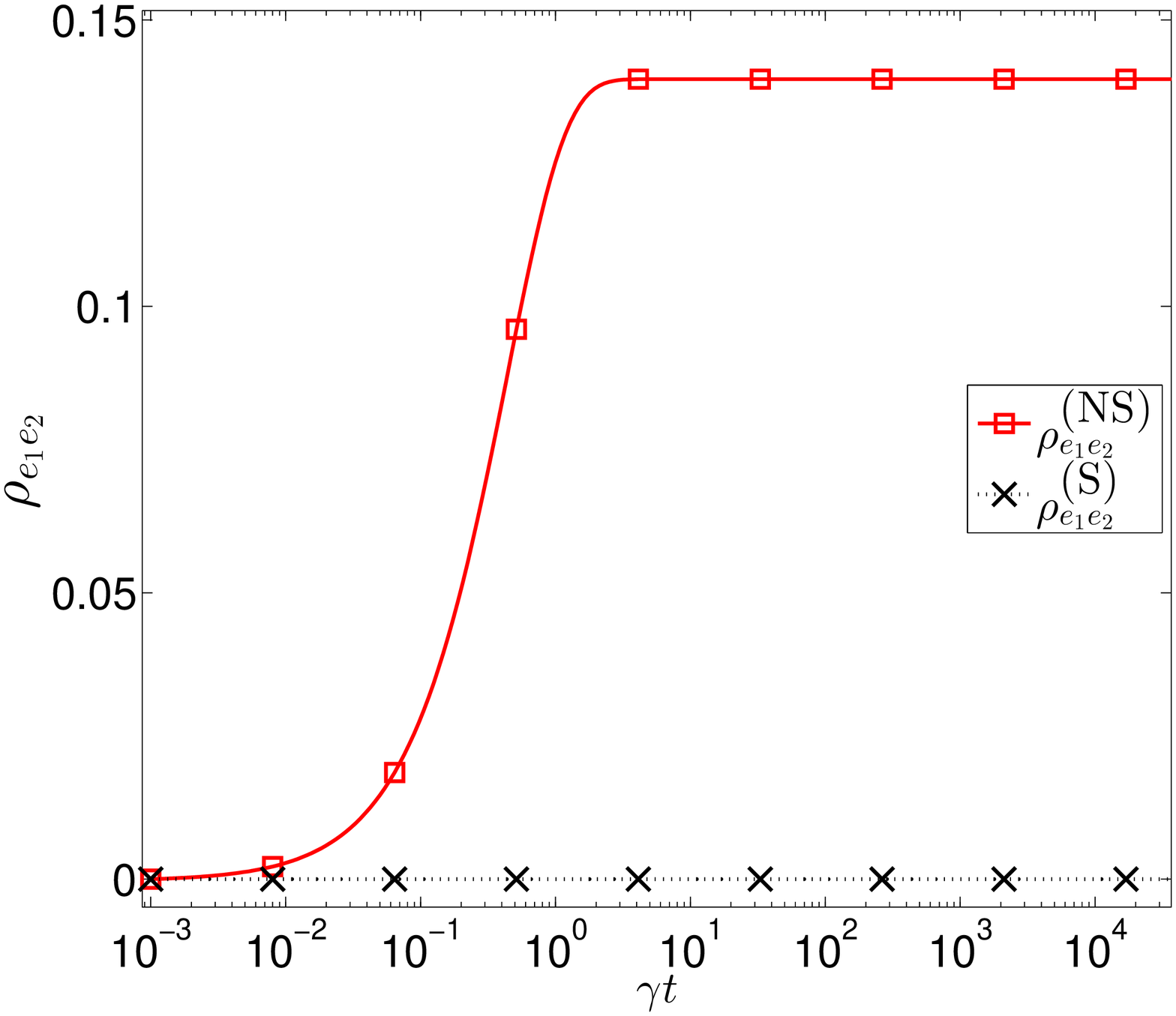}
\caption{Coherences of a Calcium atom in the small splitting regime $\Delta=0.012\gamma$ irradiated by a blackbody at $T=5800K$ whose average photon occupation number at the transition energies is $\bar{n}=0.0633$.}
\label{fig:SmallSplit}
\end{figure}

As in the case of large Zeeman splitting the non-secular population dynamics are identical to the secular population dynamics.
However, the coherences produced differ significantly; here the non-secular coherences are the same size as the populations, i.e.,
\beq
\rho_{e_1e_2}^{(\textrm{NS})}(t)=\rho_{e_ie_i}(t)~.
\label{eq:SNSCoh}
\eeq
 Hence, in the non-secular case, the stationary state is a coherent superposition.
As in the large splitting regime considered in \cref{subsec:BigDel}, the stationary coherences can be understood as the atom operating between two different baths.
\Cref{fig:SmallSplit} shows the coherences obtained from a numerical integration of the secular and non-secular Master Equations.

\begin{figure}[htp] 
\includegraphics[width=0.7\textwidth]{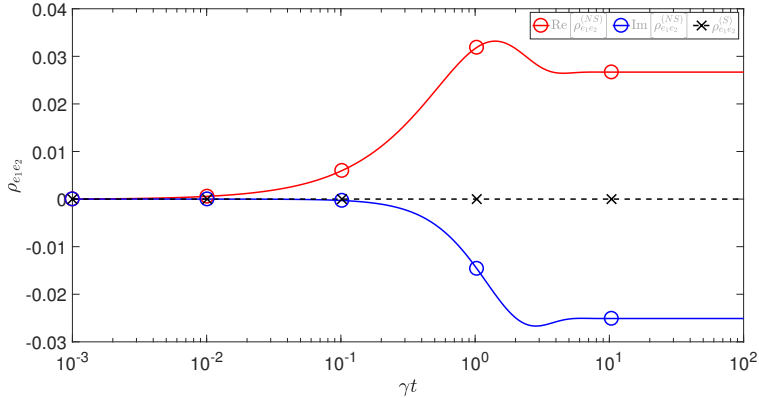}
\caption{Coherences of a Calcium atom in the intermediate splitting regime $\Delta=\gamma$ irradiated by a blackbody at $T=5800K$ whose average photon occupation number at the transition energies is $\bar{n}=0.0633$.}
\label{fig:MedSplit}
\end{figure}

In the intermediate coupling regime, an analytical solution can be obtained but will, in general, be very unwieldy.
An analytical treatment of this ``critically damped'' regime can be found in reference \cite{dodin_quantum_2016}.
However, for the purpose of this study a numerical solution in the domain of intermediate Zeeman splitting is sufficient and can be seen in \cref{fig:MedSplit}.
The secular solution in this region still shows no coherences, while the non-secular solution shows coherences betweeen those of the oscillatory coherences seen in the large splitting regime and the quasistationary coherences seen in the small splitting regime.

In summary, and as is evident in
Figs. \ref{fig:LargeSplit}
to \ref{fig:MedSplit}, the secular Master Equations show no
coherences while the non-secular equations give nonzero
coherences with dramatically different behaviors that are dependent on the
value of $\Delta/r$.

\section{Detection of Coherences by Quantum Beats}
\label{sec:Detect}
It remains to propose a measurement scheme to efficiently detect the coherences. Here
we demonstrate that the coherences are evident in emission from the irradiated atom.
To do so we compute the power spectrum of emitted radiation by using the relationship between
the electric field operators to the matter operators and, in turn, to the
density matrix elements.
The approach follows the structure in Appendix 10A of Ref. \cite{scully_quantum_1997}.

Consider the field annihilation operator $\akl(t)$ for the field mode with wavevector
$\bm{k}$ and polarization $\lambda$, in the Heisenberg picture.
For each operator define the slowly varying operator $\tilde{a}_{\bm{k}\lambda}(t)=e^{i\nu_k t}\akl(t)$ where $\akl(t)$ is the corresponding operator in the Heisenberg picture.
Let $\sigma_{ij}=\ket{j}\bra{i}$ be the atomic jump operator from state $\ket{i}$ to $\ket{j}$.
Correspondingly, define the slowly varying jump operator as $\tilde{\sigma}_{ij}(t)=e^{i\omega_{ij}t}\sigma_{ij}(t)$ where $\omega_{ij}$ is the frequency of the $\ket{i} \to \ket{j}$ transition.
Transitions between atomic energy levels are accompanied by dynamics of the field creation and
annihilation operators through the Heisenberg equations of motion. Specifically,
%

\beq
d{\tilde{{a}}_{\bm{k}\lambda}}/dt
=-i\sum_{i<j}g_{\bm{k}\lambda}^{(j,i)}\tilde{\sigma}_{ij}(t)e^{-i(\omega_{ij}-\nu_k)t}~.
\label{eq:akldiff}
\eeq

Here \Cref{eq:akldiff} assumes that the atom lies at the origin, $\tilde{\sigma}_{ij}(t)$ is the Heisenberg picture jump operator from matter state $\ket{j}$ to $\ket{i}$, $\omega_{ij}$ is the corresponding transition frequency and $\nu_k$ is the frequency of the field mode with wave vector $\bm{k}$.  No approximations have been made.
Formally integrating \cref{eq:akldiff} yields
\beq
\tilde{{a}}_{\bm{k}\lambda}(t)=\akl(0)-i\sum_{i<j}g_{\bm{k}\lambda}^{(ij)}e^{-i(\omega_ij-\nu_k)t}\int^t_0dt'\tilde{\sigma}_{ij}(t')e^{i(\omega_{ij}-\nu_k)(t-t')}~.
\label{eq:akl}
\eeq

Using the definitions
$\bm{E}^{(+)}(\bm{R},t)=\sum_{\bm{k}\lambda}E_k\bm{\epsilon}_{
\bm{k}\lambda}\akl(t)e^{i\bm{k}\cdot\bm{R}}$ and
$I(\bm{R},t)=cn_r(\epsilon_0/2)\bm{E}^{(+)}(\bm{R},t)\cdot\bm{E}^{(
-)}(\bm{R},t)$ for the
positive frequency electric field and intensity respectively,
where $n_r$ is the refractive index of the medium,
together with \cref{eq:akl} gives (with $\bm{R}$ having components $R,\theta.\phi$):
\begin{equation}
\langle I (\bm{R},t)\rangle =\frac{n_r\omega_0^4}{32\pi^2\epsilon_0c^3R^2}\left[\frac{1+\cos^2\theta}{2}\left(\rho_{e_1e_1}\left(t'\right)+\rho_{e_2e_2}\left(t'\right)\right)\\ \nonumber
+~\sin^2\theta\left(\cos2\phi\rho_{e_1e_2}^{Re}\left(t'\right)-\sin2\phi\rho_{e_1e_2}^{Re}\left(t'\right)\right)\right]
\label{eq:Intensity}
\end{equation}
for the field intensity at position $\bm{R}$ in the Calcium V-system.
Here $t' = t + R/c$.
The intensity distribution can now be integrated to obtain the response of
intensity detectors in a variety of configurations.
To maximize the detection of coherences consider three different detection set-ups:
(a) detecting the light scattered in all directions by the system, with intensity denoted $I_z$,
(b) collecting the light in the two quarter spheres with $\theta\in[0,\pi]$
and $\phi \in[-\pi/4,\pi/4]\cup[3\pi/4,5\pi/4]$, with intensity denoted $I_A$;
(c) light collected in the two quarter spheres with $\theta\in[0,\pi]$ and $\phi \in[0,\pi/2]\cup[\pi,3\pi/2]$ and denoted $I_B$.
These three observables are directly related to the density matrix elements as follows:
\beqs
\beq
I_z=\frac{8\pi}{3}I_0(\rho_{e_1e_1}+\rho_{e_2e_2})
\eeq
\beq
I_A=\half I_z+\frac{8}{3}I_0\rho_{e_1e_2}^{Re}
\eeq
\beq
I_B=\half I_z-\frac{8}{3}I_0\rho_{e_1e_2}^I
\eeq
\label{eq:Det1}
\eeqs
where $I_0=n\omega_0^4/32\pi^2\epsilon_0c^3r^2$.

Alternatively, also consider the "complementary wedges" to those used for $I_A$ and $I_B$ and
denote these intensities $I_A'$ and $I_B'$ (e.g. $I_A'$ can be obtained
from $\theta\in[0,\pi]$ and $\phi \in[\pi/4,3\pi/4]\cup[5\pi/4,7\pi/4]$).
The coherences can then be directly extracted from these intensities as:
\beqs
\beq
I_A-I_A'=\frac{16}{3}I_0\rho_{e_1e_2}^{Re}
\eeq
\beq
I_B'-I_B=\frac{16}{3}I_0\rho_{e_1e_2}^I
\eeq
\label{eq:Det2}
\eeqs
The detection schemes outlined in \cref{eq:Det1,eq:Det2}
maximize the strength of the detected quantum beat signal as
they collect light from all regions where the interference
effects are of the same sign. These results hold for both
$\Delta/r$ regions and give signals proportional to the
coherences that are   dramatically different for the secular and non-secular cases, as
shown in Figs.  \ref{fig:LargeSplit} to
\ref{fig:MedSplit}. The intensity of the signals is given by
[$2\rho_{e_1e_2}/\pi(\rho_{e_1e_1}+\rho_{e_2e_2})]I_z$ where
$I_z$ is the total fluorescence intensity given by
\cref{eq:Det1}.

\section{Conclusion}
\label{sec:Concl}

We have proposed an experimental procedure for distinguishing
between secular and non-secular dynamics in excitation with
incoherent light, using Calcium atoms in a stationary magnetic
field interacting with a
polarized  beam  of  incoherent light. Specifically, angle-resolved fluorescence measurements allow for the detection of coherences
between  the  excited  states, displaying significantly different
behavior for the non-secular and secular cases. The experimental
results will provide deep insights into the fundamental question
of the validity of secular vs non-secular Master Equations, afford a direct method for observing Fano coherences
along with the opportunity to observe stationary coherences arising from the coupling
of the Ca atom to two different photon baths.

{\bf Acknowledgements}: PB acknowledges research support from
the U.S. Air Force of Scientific Research
under contract number FA9550-13-1-0005.  AV acknowledges support from a Society in Science Branco Weiss Fellowship.

\bibliography{thesis}

\section{Appendix}
Section  \ref{sectIII} of the main text  provides a rigorous mathematical derivation of the secular and non-secular ME's using the Davies approach.
To highlight the relationship between this mathematical method and the typical Born-Markov and secular approximations used in the derivation of the Bloch-Redfield Equations, we present below an alternative derivation using a method similar to that of previous works that dealt with isotropic radiation\cite{tscherbul_partial_2015,tscherbul_long-lived_2014,dodin_quantum_2016,dodin_slow_2016-1}.  
This approach allows for a clear appreciation of the difference between 
excitation with a directed beam vs. that with isotropic radiation.

\subsection{Incident Directed Radiative Beam}

Consider the non-secular Bloch-Redfield Master Equations and secular (Pauli Rate Law) Equations for the Calcium system.
Here, the $4s$ to $4p_\pm$ transitions have orthogonal polarization with each transition corresponding to one direction of circularly polarized light ($\sigma^+ \perp \sigma^-$).
Furthermore, both $p_+$ and $p_-$ spontaneously decay to the ground $4s$ state with the same rate $\gamma_- =\gamma=\gamma_+$. The specific character of the system, and the use of a directed beam of incoherent light results in a different QME than previously obtained for isotropic excitation \cite{tscherbul_long-lived_2014}.
In the isotropic case:

\begin{subequations}
\begin{equation}
\dot{\rho}_{e_i e_i} =-(r_i + \gamma_i)\rho_{e_i e_i} +r_i \rho_{gg}-p(\sqrt{r_1r_2}+\sqrt{\gamma_1\gamma_2})\rho_{e_1e_2}^{Re}
\label{eq2:PQME2}%
\end{equation}
\begin{equation}
\dot{\rho}_{e_1e_2}=-\frac{1}{2}(r_1+r_2+\gamma_1+\gamma_2)\rho_{e_1e_2}-i\rho_{e_1 e_2}\Delta
 +\frac{p}{2}\sqrt{r_1 r_2}(2\rho_{gg}-\rho_{e_1
 e_1}-\rho_{e_2
 e_2})-\frac{p}{2}\sqrt{\gamma_1\gamma_2}(\rho_{e_1
 e_1}+\rho_{e_2 e_2})
\label{eq2:CQME2}%
\end{equation}
\label{eq2:2QME}%
\end{subequations}
where here and below atomic ($\hbar=1$) units are used. Here $\rho_{e_i e_j}$ are the density matrix elements in the $|e_i\rangle$ basis, overdots denote time derivatives,  $\rho^{Re}_{e_1 e_2}$ is the real part of $\rho_{e_1 e_2}$ and   $\rho^I_{e_1,e_2}$ below denotes the imaginary part.
In Eq. (\ref{eq2:2QME}), absorption and stimulated emission processes are parametrized by the incoherent pumping rates of the $\ket{g} \leftrightarrow \ket{e_i}$ transitions $r_i= B_iW(\omega_{g e_i})$, given by the product of the
Einstein $B$-coefficients $B_i=\pi |\mu_{g e_i}|^2/(3\epsilon_0)$ and the intensity of the incident spectrally broad radiation at the corresponding transition frequencies $W(\omega_{g e_i})$.
Spontaneous emission processes are governed by the radiative decay widths of the excited states $\gamma_i=\omega_{g e_i}^3|\mu_{g e_i}|^2/(3\pi\epsilon_0c^3)$, $\Delta=\omega_{e_1 e_2}$ gives the excited state energy splitting, and $p=\bm{\mu}_{g e_1}\cdot\bm{\mu}_{g e_2}/(|\mu_{g e_1}| |\mu_{g e_2}|)$ quantifies the alignment of the $\ket{g} \leftrightarrow \ket{e_i}$ transition dipole moments, $\bm{\mu}_{g e_i}$.
The radiative decay widths and incoherent pumping rates are related by the thermal occupation number, $\nbar$ of the electromagnetic field such that $r_i=\nbar\gamma_i$.

Consider now excitation with a polarized beam of light derived from a spectrally broadened laser source. If the $W(\omega_{ge_i})$  is flat over the two $|g\rangle \rightarrow |e_i\rangle$ transitions, then the above equations provide a useful starting point for the case of interest. However, unlike a blackbody source that
typically produces unpolarized isotropic radiation with a thermal occupancy that depends only on the magnitude of the wave vector (or equivalently the frequency) $\left< {n}_{\bm{k},\lambda}\right>=\nbar_k$,
a polarized beam of spectrally broadened light shows the same frequency dependence in the
neighborhood of the transitions of interest,
but has a strong dependence on the wavevector direction and polarization mode.
Defining the beam propagation direction as $\hat{z}$ and considering $\hat{x}$ polarized light gives:
\begin{equation}
\left<{n}_{\bm{k},\lambda}\right>=\delta_{\bm{\hat{k}},\bm{\hat{z}}}\delta_{\bm{\hat{\epsilon}}_{\bm{\hat{k}},\lambda},\bm{\hat{x}}}\nbar_k
\label{eq:nbar}
\end{equation}
where $\bm{\hat\epsilon}_{\bm\hat{k},\lambda}$ is the polarization vector for the field mode with wave vector $\bm{k}$ and polarization $\lambda$.
Thus, the incoherent pumping shows directional dependence, whereas the spontaneous emission terms are isotropic and are unchanged from the isotropic case previously considered \cite{dodin_quantum_2016,tscherbul_partial_2015,tscherbul_long-lived_2014}.
Using the same perturbative expansion that provided the Master Equations for the V-system excited by incoherent isotropic radiation \cite{tscherbul_partial_2015}~gives the  light-matter coupling coefficients for the polarized directed field driven transitions,
with the following replacement of Eq. (6) in Ref \cite{tscherbul_partial_2015}:
\begin{equation}
\sum_{\bm{k}\lambda}\sum_{ij}\sum_{lm}g_{\bm{k}\lambda}^{(i,j)}g_{\bm{k}\lambda}^{(l,m)}\langle\akl^\dagger\akl\rangle\to\sum_{\bm{k}\lambda}\sum_{m,m'=\pm}g_{\bm{k}\lambda}^{(g,e_m)}g_{\bm{k}\lambda}^{(g,e_{m'})}\langle\nbar_{\bm{k}\lambda}\rangle \equiv \textrm{C}
\label{eq:Coeff}
\end{equation}
where $g_{\bm{k}\lambda}^{(\pm)}=[(h \nu_{\bm{k}})/(2 \epsilon_0 V)]^{1/2}\bm{\mu}_{\pm}\cdot\bm{\hat{\epsilon}}_{\bm{k}\lambda}$ is the matter-field coupling coefficient for the $\ket{4s}\leftrightarrow\ket{4p_\pm}$ transition and $\bm{\mu}_{\pm}$ is the transition dipole moment of the corresponding transition.
The indices $m$ and $m'$ denote a sum over the magnetic quantum numbers of the $4P$ states comprising the excited state manifold.
For convenience, we have denoted the right side of Eq.
(\ref{eq:Coeff}) as C.

Substituting \cref{eq:nbar} into \cref{eq:Coeff} and evaluating the angular summation over $\bm{k}$ and the polarization sum over $\lambda=1,2$ reduces Eq. (\ref{eq:Coeff}) to
\begin{equation}
\textrm{C} = \sum_k \mu_{g,e_m}\mu_{g,e_{m'}}(\bm{\hat{\mu}}_{g,e_m}\cdot\bm{\hat{x}})(\bm{\hat{\mu}}_{g,e_{m'}}\cdot\bm{\hat{x}})\nbar_k
\label{eq:Coeff2}
\end{equation}
where $\mu_{g,e_m}=|\bm{\mu}_{g,e_m}|$ is the magnitude of the transition dipole moment and $\bm{\hat{\mu}}_{g,e_m}=\bm{\mu}_{g,e_m}/\mu_{g,e_m}$ is the unit vector in the direction of the transition dipole moment.
To determine the $\hat{x}$ component of the transition dipole moments recall that $\ket{4p_{m=\pm1}}=1/\sqrt{2}(\ket{4p_x}\pm i\ket{4p_y})$ and that $\bra{4p_x}\mu\ket{4p_y}=0$ by angular momentum selection rules.
Furthermore, $\bra{4s}\mu\ket{4p_x}$ is parallel to $\hat{x}$ and $\bra{4s}\mu\ket{4p_y}$ is parallel to $\hat{y}$ by the symmetry of the p orbitals
so that $\bm{\hat{\mu}}_{g,e_m}\cdot\bm{\hat{x}}=1/\sqrt{2}$.
Substituting into \cref{eq:Coeff2} allows for the evaluation of the coefficients in the perturbative expansion as

\begin{equation}
\textrm{C} = \sum_k\frac{\mu_{g,e_m}\mu_{g,e_{m'}}}{2}\nbar.
\label{eq:Coeff3}
\end{equation}

Noting that $\mu_{g,e_1} \perp \mu_{g,e_{-1}}$ we get $p=0$ for the spontaneous emission terms.
Therefore, the alignment parameters for the isotropic spontaneous emission terms and for the directional pumping terms
are different.
Further, note that when $m=m'$ in \cref{eq:Coeff3}  the incoherent pumping rates are given by
\begin{equation}
r_{\pm1}=\frac{\gamma_{\pm1}}{4}\nbar_k=\frac{\gamma}{4}\nbar_k=r.
\label{eq:r}
\end{equation}
Here we have used $\gamma_{+1}=\gamma_{-1} = \gamma$ and redefined the pumping rates by analogy to the isotropic case but attenuated by a factor of four since the only radiation available is that parallel to the beam direction.
Using these definitions for the case $m \neq m'$ in \cref{eq:Coeff3} gives the coefficients for the non-secular terms that couple the populations and the coherences.

\subsubsection{Non-secular Master Equation}
Combining these properties within the non-secular Master Equations for the isotropic case gives the following completely positive non-secular Master Equations:
\begin{subequations}\label{eq:ANSQME}%
\begin{equation}\label{eq:ANSPop}%
\dot{\rho}_{e_ie_i}=r\rho_{gg}-(\gamma+r)\rho_{e_ie_i}-r\rho_{e_1e_2}^{Re}
\end{equation}
\begin{equation}\label{eq:ANSCR}%
\dot{\rho}_{e_1e_2}^{Re}=r\rho_{gg}-(\gamma+r)\rho_{e_1e_2}^{Re}+\Delta\rho_{e_1e_2}^I -\frac{r}{2}\left(\rho_{e_1e_1}+\rho_{e_2e_2}\right)
\end{equation}
\begin{equation}\label{eq:ANSCI}%
\dot{\rho}_{e_1e_2}^I=-(\gamma+r)\rho_{e_1e_2}^I -\Delta\rho_{e_1e_2}^{Re}
\end{equation}
\end{subequations}
where $\Delta=\mu_BB$ is the Zeeman shift between the $4p_\pm$ states. Here, the states have been labelled as $\ket{e_1}=\ket{4p_-}$, $\ket{e_2}=\ket{4p_+}$ and $\ket{g}=\ket{4s}$.
As noted in the text, these equations are the same as those for excitation with
isotropic radiation  [Eq. (\ref{eq2:2QME}) with $p=1, \gamma_1=\gamma_2=\gamma, r_1=r_2=r$]
with $(\gamma + r)$ replaced by $r$ in the last two terms in Eqs. (\ref{eq2:PQME2}) and (\ref{eq2:CQME2}).

\subsubsection{Secular Rate-Law Equations}
\label{subsec:Sec}
The secular approximation neglects the non-secular terms that couple the populations and coherences in \cref{eq:ANSQME} to give the following secular Master Equations:
\begin{subequations}\label{eq:ASQME}%
\begin{equation}\label{eq:ASPop}%
\dot{\rho}_{e_ie_i}=r\rho_{gg}-(\gamma+r)\rho_{e_ie_i}
\end{equation}
\begin{equation}\label{eq:ASCR}%
\dot{\rho}_{e_1e_2}^{Re}=-(\gamma+r)\rho_{e_1e_2}^{Re}+\Delta\rho_{e_1e_2}^I
\end{equation}
\begin{equation}\label{eq:ASCI}%
\dot{\rho}_{e_1e_2}^I=-(\gamma+r)\rho_{e_1e_2}^I -\Delta\rho_{e_1e_2}^{Re}
\end{equation}
\end{subequations}
As a consequence of the decoupling of the coherences from the populations, a system initially in an incoherent mixture of states will not develop coherences between any of its states and hence will remain an incoherent mixture.
This contrasts with \cref{eq:ANSQME} where, for example, a system initially in the ground state will generate coherences between the excited states.
 In the case of \cref{eq:ANSQME} these coherences would result in the localization of a $p$ orbital along a given axis in the $xy$ plane, with the phase of the coherences defining the axis.

%

\end{document}